\documentclass[12pt]{article}
\usepackage{graphicx}
\usepackage{latexsym}
\usepackage{amssymb}

\newcommand{\dd}{\mbox{\rm d}}

\newcommand{\gam}{\gamma}

\newcommand{\ul}{\underline}

\newcommand{\be}{\begin{equation}}
\newcommand{\bear}{\begin{eqnarray}}
\newcommand{\ear}{\end{eqnarray}}
\newcommand{\ee}{\end{equation}}
\newcommand{\lbl}{\label}
\newcommand{\ci}{\cite}

\newcommand{\vs}{\vspace}

\begin{document}

\baselineskip .5cm

\thispagestyle{empty}

\ 

\vs{1.5cm}

\begin{center}

{\bf \LARGE Towards the Unification of Gravity and other Interactions:
What has been Missed?}

\vs{6mm}

M. Pav\v si\v c

Jo\v zef Stefan Institute, Jamova 39, SI-1000 Ljubljana, Slovenia;\\
E-mail: matej.pavsic@ijs.si

\vs{1.5cm}

ABSTRACT
\end{center}

Faced with the persisting problem of the
unification of gravity with other fundamental
interactions we investigate the possibility of a
new paradigm, according to which the basic space
of physics is a multidimensional space ${\cal C}$
associated with matter configurations. We consider
general relativity in ${\cal C}$. In spacetime, which is a
4-dimensional subspace of ${\cal C}$, we have not only the
4-dimensional gravity, but also other interactions,
just as in Kaluza-Klein theories. We then consider
a finite dimensional description of extended
objects in terms of the center of mass, area, and
volume degrees of freedom, which altogether form a
16-dimensional manifold whose tangent space at any
point is Clifford algebra Cl(1,3). The latter
algebra is very promising for the unification, and
it provides description of fermions.

\vs{8mm}

\section{Introduction}

Our current physical theories are mostly formulated in terms of spacetime,
a 4-dimensional manifold whose points correspond to
`events' associated with, e.g., collisions of particles.
General relativity is based on the concept of curved
spacetime which enables description of gravity. One possible
generalization is to consider higher dimensional spacetimes which,
besides 4D gravity, include other fundamental interactions.
Another possible generalization is to assume that fundamental
objects are not point particles but strings, whose excitations
contain fundamental particles and interactions. Though very
promising, those theories have encountered numerous difficulties,
and currently we still do not have a generally accepted `unified
theory'.

When considering extended objects,
there is a possibility of formulating the theory in terms
of the corresponding configuration space. Such idea has been
considered before in numerous illuminating works by
Barbour \ci{Barbour}. For some other approaches see refs.\,
\ci{AndersonConfig}. Here I will adopt the view that general
relativity should be extended to configuration space ${\cal C}$, a
multidimensional manifold equipped with metric, connection and
curvature. In this picture a 4-dimensional spacetime is just
a subspace of ${\cal C}$, associated with the degrees
of freedom of a single point particle. In order to describe
not only 4D gravity, but also other interactions, we can
employ the extra dimensions of ${\cal C}$.

\section{Generalizing relativity}
\subsection{Configuration space replaces spacetime}

Let us consider a system of point particles whose coordinates
are $X_i^\mu$, where $\mu = 0,1,2,3$,  $i = 1,2,..., N$, $N$ being
the number of particles in the configuration. We can consider
$X_i^\mu$ as coordinates in a $4 \times N$-dimensional configuration
space ${\cal C}$, therefore it is convenient to introduce a more
compact notation:
\be
 \dot X_i ^\mu  \, \equiv \dot X^{(i\mu )} \, \equiv \,\dot X^M 
 \,,\,\,\,\,\,\,\,M\, = \,(i\mu )
 \lbl{1.1}
\ee
Let us assume that the action for such system is
\be
I[X^M ]\, = \,M\int {{\rm{d}}\tau } \,\,[\dot X^M \dot X^N G_{MN} (X^M )]^{1/2}
\lbl{1.2}
\ee
which is proportional to the length of a worldline in ${\cal C}$.
Constant\footnote{Symbol $M$, of course, has a different meaning when
occurring as an index.}
 $M$ has the role of {\it mass} in ${\cal C}$, whereas $\tau$
is an arbitrary monotonically increasing parameter.

An action which is equivalent to (\ref{1.2}) is the Schild action
\be
I[X^M ]\, = \,\,\int {{\rm{d}}\tau } \,\,\dot X^M \dot X^N \,\frac{M}{K}
 \, G_{MN} (X^M )
\lbl{1.3}
\ee
From the equations of motion derived from (\ref{1.3}) it follows
that
\be
{\dot X}^M {\dot X}^M G_{MN} = K^2
\lbl{1.3a}
\ee
Therefore (\ref{1.3})
is a gauge fixed form of the action (\ref{1.2}).

We will assume that configuration space ${\cal C}$ is a manifold
equipped with metric $G_{MN}$, connection and curvature (that in
general does not vanish).

In particular, for the block diagonal metric
 \be
G_{MN} \, \equiv \,G_{(i\mu )(j\nu )}  = \,g_{(i\mu )(j\nu )}  = \,\left( {\begin{array}{*{20}c}
   {g_{\mu \nu } (x_1 )} & 0 & 0 & {...}  \\
   0 & {g_{\mu \nu } (x_2 )} & 0 & {...}  \\
   0 & 0 & {g_{\mu \nu } (x_3 )} & {...}  \\
    \vdots  &  \vdots  &  \vdots  &  \vdots   \\
\end{array}} \right)
\lbl{1.4}
\ee
the action (\ref{1.3}) becomes
\be
I[\dot X_i ^\mu  ]\, = \int {{\rm{d}}\tau } \sum\limits_i^{} 
{\dot X_i ^\mu  \dot X_i ^\nu  \,\frac{{M}}{{K}}\,g_{\mu \nu } (X_i ^\mu  )}
\lbl{1.5} 
\ee
Rewriting eq.\,(\ref{1.3a}) as
\be
     \frac{{\dot X}_1^2}{K^2} = 1 - \frac{{\dot X}_2^2}{K^2}  -  \frac{{\dot X}_3^2}{K^2}
     - ... - \frac{{\dot X}_N^2}{K^2}
\lbl{1.6}
\ee
and introducing a new constant
$m_1^2 \equiv p_1^2 = M^2 - p_2^2 - p_3^2 - ... - p_N^2$, we find
\be
         \frac{M^2}{K^2}=\frac{m_1^2}{k_1^2}  ~~~~~~~ {\rm or} ~~~ 
         \frac{M}{K}=\frac{m_1}{k_1}
\lbl{1.7}
\ee
Here $k_i^2 = X_i^2 = g_{\mu \nu} \dot X_i ^\mu \dot X_i ^\nu$, and
$p_i^2 \equiv p_i^\mu p_i^\nu g_{\mu \nu}$. Since the above derivation
can be repeated for any $i=1,2,...,N$, we have
\be
\frac{M}{K} = \frac{m_i}{k_i}
\lbl{1.8}
\ee
Inserting the latter relation into the action (\ref{1.5}) we
obtain the Schild action for a system of relativistic particles
moving in a gravitational field $g_{\mu \nu}$, in fact the sum
of Schild action for individual particles. Thus the ordinary
relativistic theory for a many particle system is just a special
case of the general action (\ref{1.2}) or (\ref{1.3}) for
a particular, block diagonal metric (\ref{1.4}).

By allowing for a more general metric, that cannot be transformed into the
form (\ref{1.4}) by a choice of coordinates in ${\cal C}$,
we go beyond the ordinary theory.

Configuration space ${\cal C}$ is the space of possible ``instantaneous" 
configurations in $M_4$. Its points are described by coordinates 
$x^M \equiv x_i^\mu$. A given configuration traces a worldline 
$x^M = X^M (\tau)$ in ${\cal C}$.

A dynamically possible worldline in ${\cal C}$ is a geodesic in ${\cal C}$,
and it satisfies the variation principle based on the action (\ref{1.2}).

Instead of considering a fixed metric of ${\cal C}$, let us assume that the
metric $G_{MN}$ is dynamical, so that the total action contains a kinetic term
for $G_{MN}$:
\be
I[X^M ,G_{MN} ] = \,I_m  + I_g 
\lbl{1.9}
\ee
where
\be
I_m \, = \int {{\rm{d}}\tau } \,\,M\,(G_{MN} \,\dot X^M \dot X^N )^{(1/2)} \, = 
\,\int {{\rm{d}}\tau } \,\,M\,(G_{MN} \,\dot X^M \dot X^N )^{(1/2)} 
\,\delta ^D \,(x\, - \,X(\tau ))\,{\rm{d}}^D x
\lbl{1.10}
\ee
and
\be
I_g \, = \,\frac{1}{{16\pi G_D }}\int {d^D x\,\sqrt {|G|} } \,{\cal R}
\lbl{1.10a}
\ee
Here ${\cal R}$ is the curvature scalar in ${\cal C}$.
So we have {\it general relativity in configuration space} ${\cal C}$.
We have arrived at a theory which is analogous to Kaluza-Klein theory.
Configuration space is a higher dimensional space, whereas
spacetime $M_4$ is a 4-dimensional subspace of ${\cal C}$,
associated with a chosen particle.

The concept of configuration space can be used either in 
macrophysics or in microphysics. Configuration space
associated with a system of point particles is finite
dimensional. Later we will discuss infinite dimensional
configuration spaces associated with strings and branes.

\subsection{Equations of motion for a configuration
of point particles}

The equations of motion derived from the action (\ref{1.9})
are the Einstein equations in configuration space ${\cal C}$.
Let us now split the coordinates of ${\cal C}$ into
4-coordinates $X^\mu \equiv X^{1 \mu}~, ~~~~\mu=0,1,2,3$
associated with  position of a chosen particle, labeled by 1,
and the remaining coordinates $X^{\bar M}$:
\be
X^M \, = \,(X^\mu  ,\,X^{\bar M} )\,
\lbl{1.11}
\ee
The quadratic form occurring in the action (\ref{1.2}) can then
be split---according to the well known procedure of Kaluza-Klein
theories---into a 4-dimensional part plus the part due to the
extra dimensions of configuration space ${\cal C}$:
\be
\dot X^M \dot X^N \,G_{MN} \, = \,\dot X^\mu  \dot X^\nu  
g_{\mu \nu } \, + \,{\rm{extra}}\,{\rm{terms}}
\lbl{1.12}
\ee

More precisely, if for the metric of ${\cal C}$ we take the ansatz
\be
G_{MN} \, = \,\left( \begin{array}{l}
 g_{\mu \nu }  + A_\mu  ^{\bar M} A_\nu  ^{\bar N} \phi _{\bar M\bar N} 
 \,,\,\,\,\,\,A_\mu  ^{\bar N} \phi _{\bar M\bar N}  \\ 
 \,\,\,\,\,\,\,\,\,A_\nu  ^{\bar N} \phi _{\bar M\bar N} 
 \,,\,\,\,\,\,\,\,\,\,\,\,\,\,\,\,\,\,\,\,\,\,\,\,
 \phi _{\bar M\bar N} \,\,\,\,\,\,\,\,\,\,\, \\ 
 \end{array} \right)
\lbl{1.13}
\ee
then we obtain
\be
\dot X^M \dot X^N \,G_{MN} \, = \,\dot X^\mu  \dot X^\nu  g_{\mu \nu } \, + 
\,\dot X_{\bar M} \dot X_{\bar N} \,\phi ^{\bar M\bar N}
\lbl{1.14}
\ee
where
\be
\dot X_{\bar M} \, = \,G_{\bar MN} \dot X^N \, = \,A_{\bar M\mu } \dot X^\mu  
 + \phi _{\bar M\bar N} \dot X^{\bar N}
\lbl{1.15}
\ee

Inserting expression (\ref{1.14}) into the action (\ref{1.10}), we have
\be
I_m [X^\mu  ,X^{\bar M}] \, = \,M\int {d\tau \,\left[ {\dot X^\mu  \dot X^\nu  g_{\mu \nu }  + 
\,\phi ^{\bar M\bar N} (A_{\bar M\mu } \dot X^\mu   + \phi _{\bar M\bar J} \dot X^{\bar J} )
(A_{\bar N\nu } \dot X^\nu   + \phi _{\bar N\bar K} \dot X^{\bar K} )} \right]} ^{1/2}
\lbl{1.16}
\ee

Let us now assume that the ``internal" subspace of ${\cal C}$  admits
isometries given by the Killing vector fields $k_\alpha^{\bar J}$.
Index $\alpha$ runs over the independent Killing vectors, whereas
${\bar J}$, like ${\bar M},~{\bar N}$, runs over the ``internal" coordinates.
Then, as it is customary in Kaluza-Klein theories, we write
\be
A_\mu  ^{\bar J}  = k_\alpha  ^{\bar J} A_\mu  ^\alpha  
\lbl{1.17}
\ee
The metric $\phi^{{\bar M}{\bar N}}$ of the internal space
can be rewritten in terms of a metric
$\varphi^{\alpha \beta}$ in the space of isometries:
\be
\phi ^{\bar M\bar N}  = \varphi ^{\alpha \beta } k_\alpha  ^{\bar M} k_\beta  ^{\bar N}
+ \phi_{\rm extra}^{{\bar M}{\bar N}}
\lbl{1.18}
\ee
Here $\phi_{\rm extra}^{{\bar M}{\bar N}}$ are additional terms due
to the directions that are orthogonal to ismotries. For particular
internal spaces ${\bar {\cal C}}$, those additional terms may vanish.

The projections of momenta onto Killing vectors are
\be
p_\alpha  \, \equiv \,\,k_\alpha  ^{\bar J} P_{\bar J} 
\lbl{1.19}
\ee
We may chose a coordinate system in ${\cal C}$ in which
\be
k_\alpha  ^M  = \left( {k_\alpha  ^\mu  ,k_\alpha  ^{\bar M} } \right)\,,\,\,\,\,
k_\alpha  ^\mu   = 0,\,\,\,\,\,k_\alpha  ^{\bar M}  \ne 0
\lbl{1.20}
\ee

Variation of the action (\ref{1.16} gives\,\ci{PavsicBled}
 $$\frac{1}{\lambda }\frac{d}{{d\tau }}\left( {\frac{{\dot X^\mu  }}{\lambda }} \right) + 
 \,^{(4)} \Gamma ^\mu  _{\rho \sigma } \frac{{\dot X^\rho  \dot X^\sigma  }}{{\lambda ^2 }}
 \,\, + \,\frac{{p_\alpha  }}{m}F_{\mu \nu } ^\alpha  \,\frac{{\dot X^\nu  }}{\lambda } $$
\be
 \,\,\,\,\,\,\,\,\,\,\,\,\,\,\,\,\,\,\,\,\,\,\,\,\,\,\,\, + \,\,\,\frac{1}{{2m^2 }}
 \,\left( {\varphi ^{\alpha \beta } _{,\mu }  - \,\varphi ^{\alpha \beta } _{,\bar J}
  \,k_{\alpha '} ^{\bar J} A_\mu  ^{\alpha '} } \right) p_\alpha  p_\beta 
   \, + \,\frac{1}{{\lambda m}}\,\frac{{\dd m}}{{\dd \tau }} = \,0 \\ 
\lbl{1.21}
\ee
where
$\lambda = \left( {\dot X^\mu  \dot X^\nu  \,g_{\mu \nu } } \right)^{1/2}$.

In the above derivation we have used the following relation between
the 4-dimensional and the higher dimensional mass
\be
\frac{m}{M}\, = \,\,\left( {\frac{{\dot X^\mu  X^\nu  
g_{\mu \nu } }}{{\dot X^M \dot X^N \,G_{MN} }}} \right)^{1/2} 
\lbl{1.22}
\ee
which, for the special case of the block diagonal metric $G_{MN}$
has been already given in eq.\,(\ref{1.7}).

From eq.\,(\ref{1.21}), in which $p_\alpha$ have the role of gauge charges, we see
that $m$ has the role of {\it inertial mass} in 4-dimensions. The 4-dimensional mass $m$
is given by higher dimensional mass $M$ and the contribution due to the extra
components of momentum $P_{\bar M}$:
\be
m^2 \, = \,g^{\mu \nu } p_\mu  p_\nu  \, = \,\,M^2 \, - 
\,\,\phi ^{\bar M\bar N} p_{\bar M} p_{\bar N} \, =
   M^2 - \varphi ^{\alpha \beta } \,p_\alpha  p_\beta  
\lbl{1.23}
\ee
These extra components $P_{\bar M}$ are in fact momenta of all other particles
within the configuration. In general $m$ is not constant, but in configuration spaces
with suitable isometries it may be constant. In the last step in eq.\,(\ref{1.23}),
for simplicity, we have considered the case in which the terms with
$\phi_{\rm extra}^{{\bar M}{\bar N}}$, occurring in eq.\,(\ref{1.18}), vanish.

A configuration under consideration can be the universe. Then, according to this theory,
the motion of a subsystem, approximated as a point particle, obeys the law of motion
(\ref{1.21}). Besides the usual 4-dimensional gravity, there are extra forces. 
They come from the generalized metric, i.e., the metric of configuration space. Since
the inertial mass of a given particle
depends on momenta of other particles and their states of motion (their momenta),
the Mach principle is automatically incorporated in this theory.
Such approach opens a Pandora's box of possibilities to revise our current
views on the universe. Persisting problems, such as the horizon problem,
dark matter, dark energy, the Pioneer effect, etc., can be examined afresh  within
this theoretical framework based on the concept of configuration space.

Locality, as we know it in the usual 4-dimensional relativity, no longer
holds in this new theory, at least not in general. But in particular,
when the metric of ${\cal C}$ assumes the block diagonal form (\ref{1.4}),
we recover the ordinary relativity (special and general), together
with locality. However, it is reasonable to expect that metric
(\ref{1.4}) may not be a solution of the Einstein equations in ${\cal C}$.
Then the ordinary relativity, i.e., the relativity in $M_4$, could
be recovered as an approximation only. Even before going
into the intricate work of solving the equations of general relativity
in ${\cal C}$, we already have a crucial prediction, namely that
locality in spacetime holds only approximately. When considering
the universe within this theory, we have to bear in mind that
the concept of spacetime has to be replaced by the concept of
configuration space ${\cal C}$. Locality in $M_4$ has thus to
be replaced by locality in ${\cal C}$. More technically this means
that, instead of differential equations in $M_4$ (e.g., the Einstein equations),
we have differential equations in ${\cal C}$: a given configuration
(a point in ${\cal C}$) can only influence a nearby configuration
(a nearby point in ${\cal C}$). Only in certain special cases
this translates into the usual notion of locality in $M_4$
(a subspace of ${\cal C}$). The so called `horizon problem'
does not arise in this theory.

\section{Configuration spaces associated with strings and branes}

Instead of a system of point particles we can consider extended
objects such as strings and, in general, branes. A brane configuration
is described by the set of functions $X^\mu (\xi^a)$, where $\xi^a$,
$a=1,2,...,n$ is a set of parameters on the brane.
We will consider a brane configuration as a point in an infinite
dimensional configuration space, called brane space ${\cal M}$.
Following refs.\,\ci{PavsicBook,PavsicPortoroz,PavsicBled}, we will therefore
use a condensed notation
\be
X^\mu  (\xi ^a )\, \equiv \,X^{\mu (\xi )} \, \equiv X^M 
\lbl{2.1}
\ee
We assume that the branes within classes of tangentially deformed branes 
are in principle physically distinct objects. 
All such objects are represented by different points of 
${\cal M}$-space.

Instead of one brane we can take a 1-parameter family of
branes $ X^\mu  (\tau,\xi ^a )\, \equiv \,X^{\mu (\xi )}(\tau)
 \, \equiv X^M (\tau) $, i.e., a curve (trajectory) in ${\cal M}$.
In principle every trajectory is kinematically possible.
A particular dynamical theory then selects which amongst
those kinematically possible branes and trajectories are
dynamically possible. We assume that dynamically possible
trajectories are {\it geodesics} in ${\cal M}$ determined by
the minimal length action \ci{PavsicBook,PavsicPortoroz}:
\be
I[X^M ] = \int {{\rm{d}}\tau } \,\,(\rho _{MN} \,\dot X^M \dot X^N )^{(1/2)} 
\lbl{2.2}
\ee
Here $\rho_{MN}$ is the metric of ${\cal M}$.

In particular, if metric is
\be
\rho _{MN} \, \equiv \,\rho _{\mu (\xi ')\nu (\xi '')} \, = 
\,\kappa \,\frac{{\sqrt {|f(\xi ')|} }}{{\sqrt {\dot X^2 \,(\xi ')} }}
\,\delta (\xi ' - \xi '')\,\eta _{\mu \nu }
\lbl{2.3}
\ee
where $f_{ab} \, \equiv \partial _a X^\mu  \partial _b X^\nu  \eta_{\mu \nu } $
is the induced metric on the brane,  $f \equiv \det \,f_{ab}$,
$\dot X^2  \equiv \dot X^\mu  \dot X^\nu  g_{\mu \nu } $, ($\eta_{\mu \nu}$
being the Minkowski metric of the embedding spacetime), then the equations
of motion derived from (\ref{3.2}) are precisely those of a Dirac-Nambu-Goto
brane \cite{PavsicBook,PavsicPortoroz}. Although we started from
a brane configuration space in which tangetially deformed branes
are considered as distinct objects, the dynamical theory based on
the action (\ref{2.2}) and the particular choice of metric (\ref{2.3})
has for solutions the branes which satisfy such constraints which imply that
only the transversal excitations are physical, whereas the tangential
excitations are nothing but reparametrizations of $\xi^a$ and $\tau$.
For more details see ref.\,\ci{PavsicBook}.

In this theory we assume that metric (\ref{2.3}) is just one particular
choice amongst many other possible metrics of ${\cal M}$. But
dynamically possible metrics are not arbitrary. We assume that
they must be solutions of the Einstein equations in ${\cal M}$
\cite{PavsicBook,PavsicPortoroz}. 

We take the brane space ${\cal M}$ as an arena for physics.
The arena itself is a part of the dynamical system,
it is not prescribed in advance.
The theory is thus background independent. It is based on the geometric
principle which has its roots in the brane space ${\cal M}$.

To sum up, the infinite dimensional brane space ${\cal M}$ has in
principle any metric that is a solution to the Einstein's equations in
${\cal M}$. For the particular diagonal metric (\ref{2.3}) we obtain the ordinary
branes, including strings. But it remains to be checked whether such
particular metric is a solution of this generalized dynamical system
at all. If not, then this would mean that the ordinary string and brane
theory is not exactly embedded into the theory based on dynamical
${\cal M}$-space. The proposed theory goes beyond that 
of the usual strings and branes. It  resolves the problem
of background independence and the geometric principle behind the
string theory . Geometric principle behind the string theory
is based on the concept of brane space ${\cal M}$, i.e., the configuration
space for branes. Occurrence of gauge and gravitational fields in
string theories is also elucidated. Such fields are due to string
configurations. They occur in the expansion of a string state
functional in terms of the Fock space basis. A novel insight is that
they occur even within the classical string theory based on the
action (\ref{2.2}) with ${\cal M}$-space metric $\rho_{MN}$,
which is dynamical and satisfies the Einstein equations in ${\cal M}$.
Multidimensionality of $\rho_{MN}$ allows for extra gauge interactions,
besides gravity. In the following we will discuss how in the infinite
dimensional space ${\cal M}$ one can factor out a finite
dimensional subspace.

\section{Finite dimensional description of extended objects}

As an example let us consider a closed string. It has infinitely many
degrees of freedom encoded in $X^\mu (\xi^a)$. But in first
approximation we can describe it by four coodinates $X^\mu$ of the
center of mass only. In the next approximation we can describe
it in terms of the coordinates $X^{\mu_1 \mu_2}$ of the oriented area
enclosed by the string. If the string has finite thickness, and thus
it actually is not a string but a 2-brane, then we can also consider
the corresponding volume degrees of freedom $X^{\mu_1 \mu_2 \mu_3}$.

In general, an extended object in 4-dimensional spacetime can be
described by 16 coordinates
\be
X^M \, \equiv \,X^{\mu _1 ...\mu _r } \,,\,\,\,\,\,\,\,r = 0,1,2,3,4
\lbl{3.1}
\ee
They denote a point in a 16-dimensional space $C$, which is a subspace
of the full infinite dimensional space ${\cal M}$, the configuration
space of the considered extended object. The oriented $r$-volumes
can be elegantly described\,\ci{PavsicArena}--\ci{CastroPavsicReview}
 by Clifford algebra\,\ci{Hestenes}. We thus consider
a space $C$ as a manifold whose tangent space at any point is Clifford
algebra. The line element in $C$ is
\be
{\rm{d}}S^2 \, = \,G_{MN} \,{\rm{d}}x^M {\rm{d}}x^N 
\lbl{3.2}
\ee
where $G_{MN}$ now denotes the metric of $C$. The latter space
is called `Clifford space'.

A worldine $x^M = X^M (\tau)$ in $C$ is associated with the motion of
the extended object, e.g., a brane.  Let the action be
\be
I = \int \dd \tau \,(G _{MN} \dot X^M \dot X^N  )^{1/2}
\lbl{3.3}
\ee
If $G_{MN}=\eta_{MN}$ is Minkowski metric, then the equations
of motion are
\be
\ddot X^M \, \equiv \,\,\frac{{\,{\rm{d}}^{\rm{2}} X^M }}{{{\rm{d}}\tau ^2 }}
\,\, = \,\,0
\lbl{3.4}
\ee

They hold for tensionless branes. For the branes with tension one has to replace
$\eta_{MN}$ with a generic metric $G_{MN}$ with non vanishing curvature.
Eq.\,(\ref{3.4}) then generalizes to the corresponding geodesic equation
\be
     \frac{1}{\sqrt{{\dot X}^2} }\, 
     \left ( \frac{{\dot X}^M}{\sqrt{{\dot X}^2}}  \right )
     + \Gamma_{JK}^M \frac{{\dot X}^J {\dot X}^K}{{\dot X}^2} = 0
\lbl{3.5}
\ee     
Such higher dimensional configuration space, associated with branes, enables
unification of fundamental interactions \` a la
Kaluza-Klein\,\ci{PavsicKaluza,PavsicMaxwellBrane}.

Functions $X^M (\tau)$ are just like functions describing the worldline
of a point particle. The four values of the index $M=\mu = 0,1,2,3$
in fact correspond to the motion of a point particle in 4-dimensional
spacetime $M_4$. The other values of the index $M = \mu _1 ...\mu _r $
for $r=0$  (a scalar), $r=2$ (bivector), $r=3$ (3-vector) and
$r=4$ (4-vector) are associated with the particle's `thickness'.
From the point of view of spacetime $M_4$ we have thus a {\it thick particle},
and not a point particle. The thickness is encoded in 16 coordinates $X^M$.
The elegance of this approach is in the fact that a thick particle in
4-dimensional spacetime $M_4$ can be described as a point particle in
16-dimensional Clifford space $C$.

Besides a point particle in Clifford space $C$ that sweeps a wordline
$X^M (\tau)$ in $C$, we can envisage a string in $C$ that sweeps
a worldsheet $X^M (\tau, \sigma)$ in $C$. The four values of the
index $M=\mu=0,1,2,3$ correspond to a string in spacetime $M_4$.
The rest of the indices $M = \mu _1 ...\mu _r$, $r=0,2,3,4$ label
those coordinates which encode the object's thickness. From the
point of view of spacetime $M_4$ the object is a {\it thick string}.
Usual strings are infinitely thin objects. Despite being called
`extended objects', they are in fact not fully extended.
Instead of infinitely thin strings we thus consider thick strings.
Their thickness is encoded in coordinates $X^M \equiv X^{\mu _1 ...\mu _r}$.

We thus have objects which are strings in a 16-dimensional space $C$.
They are described by the Nambu-Goto action, or, equivalently, by the
Polyakov action which, in the conformal gauge, takes the form
\be
     I = \frac{\kappa}{2} \int \dd \tau \, \dd \sigma ({\dot X}^M
          {\dot X}^N - X'^M X'^N) G_{MN}
\lbl{3.6}
\ee
where ${\dot X}^M \equiv \frac{\dd X^M}{\dd \tau}$ and
$X'^M \equiv \frac{\dd X^M}{\dd \sigma}$.

If the signature of the underlying space time is
$+ - - -$, i.e., $(1,3)$, then the signature of $C$ is
$(8,8)$ \ci{PavsicSaasFee}. By taking the Jackiw-Kim-Noz
definition of vacuum\,\ci{Jackiw,PavsicPseudoHarm}, one finds\,\ci{PavsicSaasFee}
that there are no central terms in the Virasoro algebra,
if the $D$-dimensional space in which the string lives has
signature $(D/2,D/2)$.

Instead of adding extra dimensions to spacetime,
we can thus start from 4-dimensional spacetime $M_4$ with signature
$(+ - - -)$ and consider the Clifford space $C$
whose dimension is 16, and signature $(8,8)$. {\it The necessary extra
dimensions for consistency of string theory are in $C$-space.} This
also automatically brings {\it spinors} into the game. It is an old
observation that spinors are the elements of left or right ideals of
Clifford algebras \ci{Teitler}--\ci{Lounesto} (see also, e.g,
refs.\,\ci{Mankoc}--\ci{Traubenberg}). In other words, spinors
are particular sort of Clifford numbers \ci{PavsicBook}.
With the string coordinates $X^M = (X,X^\mu,X^{\mu \nu},...)$ we can
associate the basis of Clifford algebra
$\gamma_M = ({\ul 1},\gam_\mu,\gam_{\mu \nu},...)$, and consider the Clifford
numbers $X^M \gam_M$ that we call `polyvectors'\footnote{
It custummary to describe position in a flat space by vectors.
If the space is a``curved" $D$-dimensional manifold, then its points
can be described in terms of vectors belonging to a vector
space ${\mathbb R}^D$ which, in particular can be the tangent space
at a chosen point of the manifold.
In the case of a curved Clifford space $C$ its points are described
in terms of `polyvectors' which are the elements of a
Clifford algebra Cl(8,8), a tangent space at a chosen point---
``an origin"---of $C$.  }.
Therefore, the string coordinate
polyvectors $X^M \gam_M$ contain spinors.
This is an alternative way of introducing spinors into the string theory
\ci{PavsicBook,PavsicSaasFee}. Attempts to achieve a deeper understanding of
the structure of supersymmetry within the context of Clifford
algebras have been undertaken in refs.\,\ci{Traubenberg,Gates}.

\section{Conclusion}

In this paper we have considered the concept of the space
of all possible matter configurations. If one assumes that
positions of all particles are fixed and only the position
of one particle is variable, then one has the space of
all possible positions of the single particle. This is just
the 4-dimensional spacetime. But the latter space is not
the whole story, since it is only a subspace of a more general
configuration space. What we have missed so far is to employ
this more general space in our theoretical considerations.
It is true that certain researchers have considered 
configuration spaces, but the idea has not yet been generally
accepted. So far we have been stuck by the fact that we
as observers, when moving around, explore only four dimensional
spacetime. Such, intuitively reasonable notion, that physical
space is associated with the degrees of freedom of a single point
particle, has to be revised. There are many particles around,
and all their degrees of freedom count; moreover,
the particles are actually not point-like, but extended.
Thus we have a multidimensional space with the prospect
for the unification of gravity with other interactions.

\vs{5mm}  

\centerline{Acknowledgement}

This work was supported by the Ministry of High Education, Science,
and Technology of Slovenia.

{\small

\end{document}